\documentclass[epjCONF]{svjour}
\usepackage{graphics}
\usepackage[varg]{txfonts} % Times fonts
\usepackage[latin1]{inputenc}
\session-title{A Universe of Dwarf Galaxies}
\begin{document}
\title{The Color-Dependent Frequency of XUV Disks In Low-Mass E/S0s}
\author{Amanda J. Moffett\inst{1}\fnmsep\thanks{\email{amoffett@physics.unc.edu}} \and Sheila J. Kannappan\inst{1} \and Andrew J. Baker\inst{2} \and Seppo Laine\inst{3} }
\institute{Dept. of Physics \& Astronomy, University of North Carolina, Phillips Hall, CB 3255, Chapel Hill, NC 27599 \and Dept. of Physics \& Astronomy, Rutgers, State University of New Jersey, 136 Frelinghuysen Rd, Piscataway, NJ 08854 \and \emph{Spitzer} Science Center, California Institute of Technology, MS 220-6, Pasadena, CA 91125}

\abstract{We identify a high frequency of Type 1 XUV disks, reflecting
recent outer disk star formation, in a sample of 31 E/S0s with stellar
masses primarily below $M$$_{*}$ $\sim$ $4 \times
10^{10}\,M_{\odot}$. Our $\sim$40\% identification rate is roughly
twice the 20\% fraction reported for late-type galaxies. Intriguingly,
in the dwarf mass regime (below \break $M$$_{*}$ $\sim$ $5 \times 10^{9}\,
M_{\odot}$) where gas fractions clearly rise, Type 1 XUV disks occur
in $\sim$70\% of \emph{red}-sequence E/S0s but only $\sim$20\% of
blue-sequence E/S0s, a population recently linked to active disk
rebuilding, especially in the dwarf regime. Our statistics are
preliminary, but could indicate that for dwarf E/S0s Type 1 XUV disks
are primarily related to weak or inefficient outer-disk star formation
rather than to star formation capable of driving substantial disk
growth. Substantial growth may instead be associated with populations
that have \emph{low} XUV-disk frequency, possibly explaining the
similar $\sim$20\% frequencies for normal late types and low-mass
blue-sequence E/S0s.}
%end of abstract
%
\maketitle
%
%\section{Introduction}
%\label{intro}
%Your text comes here. Separate text sections with
%\section{Section title}
%\label{sec:1}
%and \cite{RefJ}
%\subsection{Subsection title}
%\label{sec:2}
%as required. Don't forget to give each section
%and subsection a unique label (see Sect.~\ref{sec:1}).
%
The recent discovery of extended ultraviolet (XUV) disks (e.g.,
\cite{Th05}, \cite{Gil05}), reflecting ongoing star formation beyond
the optical radii and traditional star formation thresholds of
late-type galaxies, has provided an intriguing look at disk
growth in progress at $z$$\sim$$0$. A separate line of research on
``blue-sequence E/S0s,'' morphologically defined E/S0 galaxies on the
blue sequence in color vs. stellar mass space, may provide a direct look
at disk \emph{regrowth} in progress at $z$$\sim$$0$ (\cite{KGB}), a
process predicted by hierarchical models (e.g., \cite{SM},
\cite{Gv07}). Blue-sequence E/S0s increase dramatically in abundance
below the gas-richness threshold mass at \break$M$$_{*}$ $\sim$ $5 \times
10^{9}\, M_{\odot}$, the regime in which neutral-atomic-gas/stellar mass
ratios $\gtrsim$1 become common (\cite{K04}, \cite{KW}, see \cite{KGB}
regarding corrected mass scale). As shown in \cite{KGB} and
\cite{We10a}, many blue-sequence E/S0s display the global gas
reservoirs and specific star formation rates necessary for significant
stellar disk growth on relatively short timescales. As a result, if
the XUV-disk phenomenon is associated with disk building in general,
we might expect to observe XUV disks preferentially among
\emph{blue-sequence} E/S0s.

Although the first XUV-disk studies, such as \cite{Th07}, emphasized
late-type galaxies, XUV disks around E/S0s are increasingly being
found as well (e.g., \cite{D09}, \cite{CH}, \cite{Th10},
\cite{SR10}). However, the question remains: are XUV disks in E/S0s
related to the probable active disk-builders, the blue-sequence E/S0
population? We address this question by examining the frequency of XUV
disks in E/S0s as a function of both sequence and mass, finding a
surprising but illuminating lack of association between Type~1 XUV
disks and E/S0s at the low-mass, gas-rich end of the blue sequence.

\begin{figure}
\resizebox{0.95\columnwidth}{!}{\includegraphics{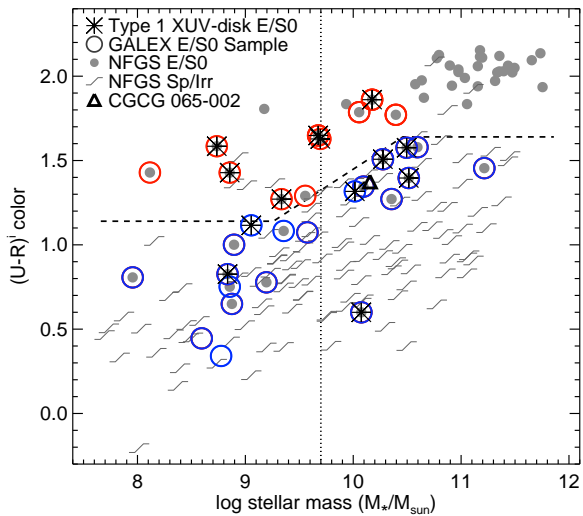} }
%\epsscale{1.}
%\plotone{Full_Redblue_lyonconf.eps}
\caption{E/S0 sample in color-stellar mass space. Small grey symbols
indicate galaxies in the NFGS, the parent sample for all but six of
our E/S0s. The dashed line divides the red and blue sequences, and the
vertical line marks the gas-richness threshold mass (\cite{KGB}). The
30 E/S0s from \emph{GALEX} program GI3-0046 are denoted by open
circles, and CGCG~065-002 from program GI5-042 is denoted by an open
triangle. Type~1 XUV-disk galaxies in this sample are marked with
asterisks.}
\label{RBfig}
\end{figure}

Our sample of 31 E/S0s encompasses all of the Nearby Field Galaxy
Survey (NFGS, \cite{Jan}) blue-sequence E/S0s and the majority of NFGS
red-sequence E/S0s with \break $M$$_{*}$~$\lesssim$~$4 \times 10^{10}\,
M_{\odot}$ (Fig.\ \ref{RBfig}), where many E/S0s have substantial gas,
and relatively undisturbed blue-sequence E/S0s with the potential for
disk regrowth are observed (\cite{KGB}). In addition to these 25 NFGS
E/S0s, we include 6 blue-sequence E/S0s from the ``HyperLeda+'' sample
of \cite{KGB}.

In \cite{Th07}, Type~1 XUV disks are defined as displaying more than
one structured UV-bright emission complex beyond a centralized
surface-brightness contour corresponding to the expected star
formation threshold (equated to an NUV surface brightness of 27.35 AB
mag\,arcsec$^{-2}$ in \cite{Th07}, roughly matching typical H$\alpha$
and HI thresholds; we label the corresponding radius $R_{\rm
\it{UVSF}}$). In addition, this definition requires that the XUV
emission take on a different morphology from any underlying optical
emission.

We classify our sample galaxies based on the Type~1 XUV-disk
definition using \emph{GALEX} NUV images with a minimum exposure time
of 1500~s. For comparison of UV and optical morphologies, we employ
DSS-II red images (http://archive.stsci.edu/dss/). These
classifications supersede the preliminary, purely visual (made without
reference to $R_{\rm \it{UVSF}}$) classifications of \cite{Me09}. The
Type~1 XUV disks in our E/S0s can extend to several times $R_{25}$ as
has been found in late types (e.g., \cite{Th05}, \cite{Gil05},
\cite{Th07}, see also \cite{ZC}). We find radial extents (to the last
measured NUV point) between $\sim$0.8 and 3 times $R_{25}$, with an
average of $\sim$1.6.

Compared to XUV disks in late-type galaxies, our E/S0 XUV disks tend
to be redder. However, using outer-disk FUV$-$NUV colors and assuming
the observed UV emission comes from young stars, we still estimate
$<$1 Gyr ages for our E/S0 XUV disks from simple stellar population
(SSP) models (although a mixed old plus young population is presumably
present, we defer multiple-component stellar population modeling to a
future paper). Comparing with UV model colors of \cite{B03} for an
instantaneous starburst with $Z=0.02$ (as in \cite{Th07} Figure 1),
our average Type~1 XUV-disk FUV$-$NUV color of $\sim$1.6 corresponds
to a stellar population with an approximate age of 500 Myr.

Although the UV upturn, i.e., UV emission associated with old stellar
populations (\cite{OC99}), is a plausible source for excess UV flux
in E/S0s, the FUV$-$$K$ colors we measure, as well as independent
evidence for actual or potential star formation (e.g., H$\alpha$ or HI
detections in all but two of our XUV-disk galaxies), largely support an
association of our XUV disks with recent star formation.

Type~1 XUV disks occur in 13/31 or 42$^{+11}_{-10}$\% of our E/S0
sample. This frequency is approximately double the 20\% Type~1
XUV-disk frequency found in \cite{Th07} for a mostly late-type galaxy
sample. We find XUV disks in both red- and blue-sequence E/S0s over a
large range in stellar mass (Fig.\ \ref{RBfig}). The widespread nature
of the XUV-disk phenome-non seems to suggest an association with
evolutionary processes affecting the galaxy population broadly, such
as gas accretion and/or minor satellite interactions.

We also observe a complex mass and sequence dependence of XUV-disk
incidence: the relative abundance of red- \emph{vs.} blue-sequence
XUV-disk galaxies \emph{seems to reverse} across the gas-richness
threshold mass. On the red sequence, the XUV-disk frequencies are low
and high (33$^{+41}_{-28}$\% and 71$^{+18}_{-26}$\%) above and below
the threshold mass, respectively. On the blue-sequence, the
corresponding frequencies are high and \emph{low} (50$\pm$20\% and
18$^{+19}_{-12}$\%). For low-mass E/S0s, the association we observe
between \emph{red} color and a high frequency of Type 1 XUV disks
suggests that these disks typically experience inefficient or
weak star formation instead of the pronounced star formation expected
on the \emph{blue} sequence in this gas-rich regime. This link between
Type~1 XUV disks and weak/inefficient star formation is supported by
previous studies (e.g., \cite{Th07}, \cite{B08}).
 
Based on these results, observing Type 1 XUV disks in E/S0s may not
provide the best indication of \emph{significant} disk
regrowth. Instead, relatively \emph{low} XUV-disk incidence, as is
found for both low-mass blue-sequence E/S0s and normal late-types
(\cite{Th07}), may be characteristic of populations with active disk
building in progress. Our findings highlight the need for a more
general UV-bright disk definition, which will be addressed in future
work.
\vspace{1.2 mm}
%\acknowledgments

We thank S. Jogee for her role in acquiring the \emph{Spitzer} data
and M. Haynes for the early release of \emph{GALEX} imaging of
NGC~3773, an archival target in our sample. We also thank L. Wei,
M. Norris, C. Clemens, and A. Leroy for useful discussions. AJM
acknowledges support from the NASA Harriett G. Jenkins Pre-doctoral
Fellowship Program. This work uses observations made with the NASA
Galaxy Evolution Explorer and \emph{Spitzer} Space
Telescope. \emph{GALEX} is operated for NASA by Caltech under NASA
contract NAS5-98034. We acknowledge support from the \emph{GALEX}
Guest Investigator program under NASA grants NNX07AT33G and
NNX09AF69G. \emph{Spitzer} is operated by the Jet Propulsion
Laboratory, Caltech under a contract with NASA. Support for this work
was also provided by NASA through an award issued by JPL/Caltech.

%\begin{table}
%\caption{Please write your table caption here.}
%\begin{tabular}{lll}
%\hline\noalign{\smallskip}
%\end{tabular}
%\end{table}
%


\begin{thebibliography}{}
%Author, Journal \textbf{Volume}, (year) page numbers
%Author, \textit{Book title} (Publisher, place year) page numbers
% etc


\bibitem{Th05}
Thilker, D., et al. 2005, ApJ, 619, L79

\bibitem{Gil05}
Gil de Paz, A., et al. 2005, ApJ, 627, L29

\bibitem{KGB}
Kannappan, S. J., Guie, J. M., \& Baker, A. J. 2009, AJ, 138, 579

\bibitem{SM}
Steinmetz, M. \& Navarro, J. F. 2002, New Astronomy, 7, 155

\bibitem{Gv07}
Governato, F., Willman, B., Mayer, L., Brooks, A., Stinson, G., Valenzuela, O., Wadsley, J., \& Quinn, T. 2007, MNRAS, 374, 1479

\bibitem{K04}
Kannappan, S. J. 2004, ApJ, 611, L89

\bibitem{KW} 
Kannappan, S. J. \& Wei, L. H. 2008, in AIP Conf. Ser. 1035, ed. R. Minchin \& E. Momjian, 163

\bibitem{We10a}
Wei, L. H., Kannappan, S. J., Vogel, S. N., \& Baker, A. J. 2010, ApJ, 708, 841

\bibitem{Th07}
Thilker, D., et al. 2007, ApJS, 173, 538

\bibitem{D09}
Donovan, J. L., et al. 2009, AJ, 137, 5037

\bibitem{CH}
Cortese, L. \& Hughes, T. M. 2009, MNRAS, 400, 1225

\bibitem{Th10}
Thilker, D., et al. 2010, ApJ, 714, L171

\bibitem{SR10}
Salim, S. \& Rich, R. M. 2010, arXiv:1004.2041

\bibitem{Jan}
Jansen, R. A., Fabricant, D., Franx, M., \& Caldwell, N. 2000a, ApJS, 126, 331

\bibitem{Me09}
Moffett, A. J., Kannappan, S. J., Laine, S., Wei, L. H., Baker, A. J., \& Impey, C. D. 2010, in ASP Conf. Ser. 423, ed. B. J. Smith, N. Bastian, S. J. U. Higdon, \& J. L. Higdon, 346

\bibitem{ZC}
Zaritsky, D. \& Christlein, D. 2007 AJ, 134, 135

\bibitem{B03}
Bruzual, G. \& Charlot, S. 2003, MNRAS, 344, 1000

\bibitem{OC99}
O'Connell, R. W. 1999, ARA\&A, 37, 603

\bibitem{B08}
Boissier, S., et al. 2008, ApJ, 681, 244


\end{thebibliography}
\end{document}